\documentclass[aps,twocolumn,prd,nofootinbib]{revtex4}
\usepackage{amsmath}
\usepackage{graphicx}
\usepackage{dcolumn}
\usepackage{bm}
\usepackage{amssymb}
\usepackage{latexsym}

\def\be{\begin{equation}}
\def\ee{\end{equation}}
\def\bea{\begin{eqnarray}}
\def\eea{\end{eqnarray}}

\begin{document}

\title{Two-field Models of Dark Energy with Equation of State Across -1
}

\author{Xiao-fei Zhang$^a$, Hong Li$^a$, Yun-Song Piao$^{a,b}$ and Xinmin Zhang$^a$}
\affiliation{${}^a$Institute of High Energy Physics, Chinese
Academy of Sciences, P.O. Box 918-4, Beijing 100049, P. R. China}
\affiliation{${}^b$Interdisciplinary Center of Theoretical
Studies, Chinese Academy of Sciences, P.O. Box 2735, Beijing
100080, China}

\begin{abstract}

 In this paper, we study the possibility of building two-field models of
dark energy
with equation of state across -1.
Specifically we will consider
two classes of models:
one consists of two scalar fields (Quintessence + Phantom) and another
includes one scalar (Phantom) and one spinor field (Neutrino).
Our studies indicate to some extent that two-field
models give rise to a simple realization of the
dynamical dark
energy model with the equation of state across $w = -1$.

\end{abstract}

\maketitle

\section{Introduction}

Recent observational data \cite{Riess, Spergel, Seljak} strongly
indicate that the Universe is spatially flat and accelerating at
the present time. In the frame of Friedmann-Robertson-Walker (FRW)
cosmology, the acceleration may be attributed to some mysterious source
called dark energy. The simplest candidate for dark
energy seems to be a small positive cosmological constant, but it
suffers from difficulties associated with the fine tuning and
coincidence problem. An alternative is a dynamical scalar field,
such as Quintessence \cite{Quin} or Phantom \cite{Caldw}. The
Phantom field violates the energy conditions, which leads to many
interesting cosmological phenomena \cite{Phantom}.

Despite the current theoretical ambiguity for the nature of dark
energy, the prosperous observational data (e.g. supernova, CMB and
large scale structure data and so on ) have opened a robust window
for testing the recent and even early behavior of dark energy
using some parameterizations for its equation of state. The recent
fits to current supernova(SN) Ia data, CMB and LSS\cite{stein}
find that even though the behavior of dark energy is to great
extent in consistency with a cosmological constant, an evolving
dark energy with the equation of state $w$ larger than -1 in the
recent past but less than -1 today is weakly favored. If such a
result holds on with the accumulation of observational data, this
would be a great challenge to the current cosmology.

The evolving dark energy with an equation of state $w$ crossing -1
during its evolution was firstly advocated and named as Quintom in
Ref. \cite{FWZ}. The Quintom models of dark energy are different
from the Quintessence or Phantom in the determination of the
evolution and fate of the universe. Generically speaking, the
Phantom model has to be more fine tuned in the early epochs to
serve as dark energy today, since its energy density increases
with expansion of the universe. Meanwhile the Quintom model can
also preserve the tracking behavior of Quintessence, where less
fine tuning is needed.
 The Quintom model with an oscillating
equation of state considered in \cite{FLPZ}can lead to the
oscillations of the Hubble constant and a recurring universe,
which in some sense unifies the early (Phantom) inflation
\cite{Piao} and current acceleration of the universe.
 This oscillating Quintom
would not lead to a big crunch nor big rip. The scale factor keeps
increasing from one period to another and leads naturally to a
highly flat universe. Since in this model the universe
recurs itself and we are
only staying among one of the epochs, the
coincidence problem in some sense is reconciled. Ref.\cite{XBZ} has
considered
 a variation of this oscillating Quintom model and studied its constraints
from SN, CMB and LSS.

There have been some efforts made on the model building of the Quintom
dark energy
based on field theory. First of all, a single scalar field with the
canonical kinetic term (Quintessence) or a negative kinetic term (Phantom)
will not be able to give rise to the equation of state across -1.
 The similar conclusion has also been obtained for the
k-essence models \cite{Vikman}.
Beyond the single scalar field theory,
Refs.\cite{FWZ} and \cite{GPZZ} have proposed an explicit model of Quintom
with two scalar fields,
 one being the
Quintessence and another being the Phantom field. This type of
model can easily lead to a scenario where at early time the
Quintessence dominates with $w
> -1$ and lately the Phantom dominates with $w<-1$,
satisfying current observations. Some recent relevant studies are
given in \cite{WHU}.

The Quintom model considered in refs.\cite{FWZ} and \cite{GPZZ}
does not include the interaction between the two scalars and is
not the most general one. In this paper, we revisit this type of
two-field Quintom model by introducing a interaction term. We will
study in detail the cosmological evolution of this model. In
ref.\cite{GPZZ} it has been shown that in the two-field Quintom
model the scaling solution dominated by the Phantom field will be
a late-time attractor in the absence of the interactions. In this
paper by a explicit calculation we will show the interactions do
not affect the Phantom-domination attractor behavior. In addition
our studies also show that this class of models will provide some
interesting possibilities of the evolution of the equation of
state which have not been considered in refs.\cite{FWZ} and
\cite{GPZZ}. Furthermore we will propose in this paper a two-field
Quintom model with a scalar and a fermion which specifically we
take to be the neutrinos and study its evolution of the equation
of state. This paper is organized as follows: in section II we
study in detail the model with one Quintessence and one Phantom;
in section III we present our model of Quintom with a Phantom and
neutrinos; section IV is our conclusion and discussions.

\section{Quintom model with Phantom field and Quintessence field}

In this section, we study a Quintom model with two scalar fields,
in which one is the Phantom field and another is the Quintessence field.
The Lagrangian for such a coupled system is given by \be {\cal
L}={\cal L}_{\phi}+{\cal L}_{\sigma}+V_{int}, \label{Ltot}\ee
where
\begin{equation}
  {\cal L}_\phi=-\frac{1}{2}\partial_\mu
  \phi\partial^\mu\phi-V(\phi) ,
  \end{equation}
  with $V(\phi)$ being the potential for the Phantom field and
\begin{equation}
  {\cal L}_\sigma=\frac{1}{2}\partial_\mu
  \sigma\partial^\mu\sigma-V(\sigma)
  \end{equation}
  with $V(\sigma)$ the potential for the Quintessence field, and
  $V_{int}$ denotes the interaction between the Phantom and the
  Quintessence fields. The equation of state of this system is
  \begin{equation} w=\frac{-\frac{1}{2}\dot \phi^{2}+\frac{1}{2}\dot \sigma^{2}-V(\phi)-V(\sigma)-V_{int} \,}
{-\frac{1}{2}\dot \phi^{2}+\frac{1}{2}\dot
\sigma^{2}+V(\phi)+V(\sigma)+V_{int} \,} .
\end{equation}

The evolution equations of the fields and the fluid for a
spatially flat FRW model are
\begin{eqnarray}
\ddot{\phi}+3H\dot{\phi}-\frac{dV(\phi)}{d\phi}-\frac{dV_{int}}{d\phi}=0 , \\
\ddot{\sigma}+3H\dot{\sigma}+\frac{dV(\sigma)}{d\sigma}+\frac{dV_{int}}{d\sigma}=0
, \\
\dot\rho_{\gamma}+3H(\rho_{\gamma}+P_{\gamma})=0 , \label{motion}
\end{eqnarray}
where $\rho_{\gamma}$
 is the density of fluid with a barotropic equation of state
 $P_\gamma =(\gamma-1)\rho_{\gamma}$ where $\gamma$ is a constant in
the range of $0<
 \gamma \leq 2$.

Now we study the cosmological evolution of the model with
lagrangian (1). The authors of ref. \cite{GPZZ} have considered a
model where $V(\phi)=V_{\phi 0}e^{-\lambda_{1}\frac{\phi}{M_{pl}}}
$, $V(\sigma)=V_{\sigma_0}e^{-\lambda_{2}\frac{\sigma}{M_{pl}}}$
and show that the late time behavior of this model is a
Phantom-domination attractor. Here we introduce an interaction
term $V_{int}=\lambda (V(\phi)V(\sigma))^{1/2}$ and study its
effect on the cosmological evolution. We follow closely the
conventional phase-plane analysis for the spatially flat FRW
models in \cite{HW} and the detailed studies on models with multi
coupled Quintessence fields in Ref. \cite{GPZ}.
 Defining firstly the variables
\bea x_{\phi}&\equiv &
\frac{\phi'}{\sqrt{6}}~,~y_{\phi}\equiv\frac{\sqrt{V(\phi)}}{\sqrt{3}H}~,~
x_{\sigma}\equiv\frac{\sigma'}{\sqrt{6}}~, \nonumber\\& &
y_{\sigma}\equiv\frac{\sqrt{V(\sigma)}}{\sqrt{3}H}~,~
z\equiv\frac{\sqrt{\rho}}{\sqrt{3}H}~, \eea the evolution Eqs.
(5-7) can be written to an autonomous system \bea x_{\phi}'&=&-
3x_{\phi}(1+x_{\phi}^{2}-x_{\sigma}^{2}-\frac{\gamma
z^{2}}{2})-\frac{\sqrt{6}}{2}\lambda_{1}y_{\phi}^{2}\nonumber\\
& & -\frac{\sqrt{6}}{4}\lambda_{1}\lambda
y_{\phi}y_{\sigma},\\
y_{\phi}'&=&  3y_{\phi}(-x_{\phi}^{2}+x_{\sigma}^{2}+\frac{\gamma
z^{2}}{2})-\frac{\sqrt{6}}{2}\lambda_{1}x_{\phi}y_{\phi},\\
x_{\sigma}'&=&-
3x_{\sigma}(1+x_{\phi}^{2}-x_{\sigma}^{2}-\frac{\gamma
z^{2}}{2})+\frac{\sqrt{6}}{2}\lambda_{2}y_{\sigma}^{2}\nonumber\\
& &+\frac{\sqrt{6}}{4}\lambda_{2}\lambda
y_{\phi}y_{\sigma},\\
y_{\sigma}'&=&
3y_{\sigma}(-x_{\phi}^{2}+x_{\sigma}^{2}+\frac{\gamma
z^{2}}{2})-\frac{\sqrt{6}}{2}\lambda_{2}x_{\sigma}y_{\sigma},\\
z'&=& 3z(-x_{\phi}^{2}+x_{\sigma}^{2}+\frac{\gamma
z^{2}}{2})-\frac{3}{2}\gamma z , \label{critic} \eea where a prime
denotes a derivative with respect to the logarithm of the scale
factor. The critical points correspond to the fixed points where
$x_{\phi}'=0$, $x_{\sigma}'=0$, $y_{\phi}'=0$, $y_{\sigma}'=0$,
$z'=0$, which have been calculated and given in Table I.

\begingroup

\begin{table*}

\begin{tabular}{c c c c c c c} \hline
Label & $x_\phi$ & $y_\phi$ & $x_\sigma$ & $y_\sigma$ & z
 & \\ \hline
$1.$ & $x_{\sigma}^2-x_{\phi}^2=1$ & 0 & & 0 & 0 & \\
$2.$ & $-\frac{\lambda_1}{\sqrt{6}}$
 & $\sqrt{ (1+\frac{\lambda_1^2}{6})}$ & 0 & 0 & 0 &   \\
$3.$ & 0 & 0 & $\frac{\lambda_2}{\sqrt{6}}$ &
 $\sqrt{ (1-\frac{\lambda_2^2}{6})}$ & 0 &   \\
$4.$ & 0 & 0 & 0 & 0 & 1 & \\
$5.$ & 0 & 0 & $\frac{3\gamma}{\sqrt{6}\lambda_2}$
 & $\sqrt{\frac{3\gamma(2-\gamma)}{2\lambda_2^2}}$
 & $\sqrt{1-\frac{3\gamma}{\lambda_2^2}}$ & \\ \hline
\end{tabular}
\caption{The list of the critical points.}

\end{table*}
\endgroup

\begingroup
\begin{table*}
\begin{tabular}{c c c c c c c} \hline
Label & $m_{1}$ & $m_{2}$ & $m_{3}$ & $m_{4}$ & Stability
 &  \\ \hline
$1.$ & $-6(1-\frac{\gamma}{2})x_{\phi c}^{2}$ &
$3(1-\frac{\sqrt{6}}{6}\lambda_{1}x_{\phi c})$
&$6(1-\frac{\gamma}{2})x_{\sigma c}^{2}$ &$ 3(1-\frac{\sqrt{6}}{6}\lambda_{2}x_{\sigma c})$ & unstable  & \\
$2.$ & $-\frac{\lambda_{1}^{2}}{2}$
 & $-\frac{1}{2}(6+\lambda_{1}^{2})$ & $-\frac{1}{2}(6+\lambda_{1}^{2})$ & $-3\gamma-
\lambda_{1}^{2}$ & stable &   \\
$3.$ & $\frac{\lambda_{2}^{2}}{2}$ &
$-3(1-\frac{\lambda_{2}^{2}}{6})$ &
$-3(1-\frac{\lambda_{2}^{2}}{6})$ &
 $-3\gamma+
\lambda_{2}^{2}$ & unstable &   \\
$4.$ & $\frac{3\gamma}{2}$ & $\frac{3\gamma}{2}$ & $-3(1-\frac{\gamma}{2})$ & $-3(1-\frac{\gamma}{2})$ & unstable &  \\
$5.$ & $\frac{3\gamma}{2}$ & $-(3-\frac{3\gamma}{2})$ &
$\frac{-3(2-\gamma)}{4}(1+\sqrt{1-\frac{8\gamma
\lambda_{2}^{2}-24\gamma^{2}}{2\lambda_{2}^{2}-\gamma
\lambda_{2}^{2}}})$
 & $\frac{-3(2-\gamma)}{4}(1-\sqrt{1-\frac{8\gamma
\lambda_{2}^{2}-24\gamma^{2}}{2\lambda_{2}^{2}-\gamma
\lambda_{2}^{2}}})$
 & unstable &  \\ \hline
\end{tabular}
\caption{The eigenvalues and stability of the critical points.}
\end{table*}
\endgroup

\begin{figure}[ht]
\begin{center}
\includegraphics[width=8cm]{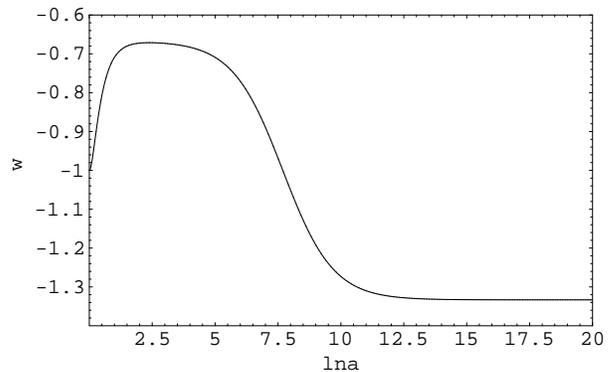}
\caption{Plot of the equation of state $w$ as a function of the scale
factor
$\ln{a}$. }
\end{center}
\end{figure}

Eqs. (9-13) can be reduced to four independent equations. To study
the stability of the critical points, we substitute the linear
perturbations about the critical points into these independent
equations and keep terms to the first-order in the perturbations. The four
perturbation equations give rise to four eigenvalues. The stability
requires the real part of all eigenvalues be negative (see
Table II for the eigenvalues of perturbation equations and the
stability of critical points).

From Table I and II, one can see that even if there exists the
interaction between the two fields, the Phantom-dominated solution
is still a late-time stable attractor. If this coupled system is
initially dominated by the Quintessence field, it will eventually
evolve into the Phantom-domination phase. Thus an equation of
state across $-1$ will be inevitable. This provides a natural
realization of the Quintom scenario. As an illustration we in Fig.
1 plot the evolution of the equation of state $w$ as a function of
the scale factor where in the numerical calculation we have taken
$V_{\phi_0}=V_{\sigma_0}= 0.35\times 10^{-46} (GeV)^4, \lambda_1 =
\lambda_2 = \lambda=1$.

Before concluding this section we point out
some interesting behavior of the equation of
the state of the two-field models.
In Fig.2
and
Fig.3, by choosing potentials and specific values of the model
parameters we show that the equation of state could be oscillated.
And the oscillations
mainly occur in Fig. 2 in the late time when redshift $z<0$, while Fig.3
give rise to one example
more interesting, in which the oscillations across $-1$ occur in
the near past, which might lead to some effects testable observationally.

\begin{figure}[htbp]
\begin{center}
\includegraphics[width=8cm]{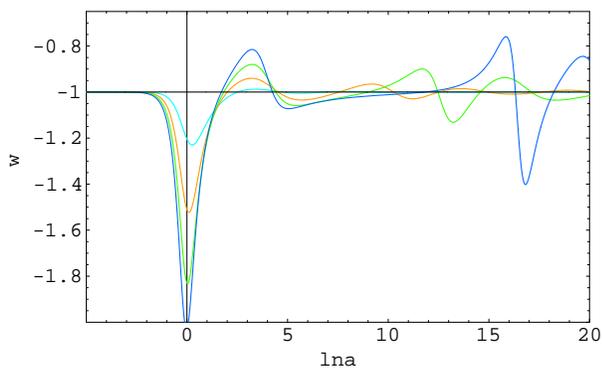}
\caption{ The state equation $w$ as a function of scale factor
$\ln{a}$ for the potential $V=
\lambda_{1}\cos{(\xi\frac{\phi}{M_{pl}})}+\lambda_{2}\cos({\alpha
\frac{\sigma}{M_{pl}}}) +\lambda \phi^{2}\sigma^{2}$, where
$\xi=0.5, \alpha = 1 , \lambda_{1}= \lambda_{2}= 2.47\times
10^{-46} GeV^4$.  The four lines  from the top to the bottom at
$\ln a=0$ correspond to $\lambda = (0.2, 0.3, 0.4, 0.5)\times
10^{-120}$ respectively.}


\end{center}
\end{figure}

\begin{figure}[htbp]
\begin{center}
\includegraphics[width=8cm]{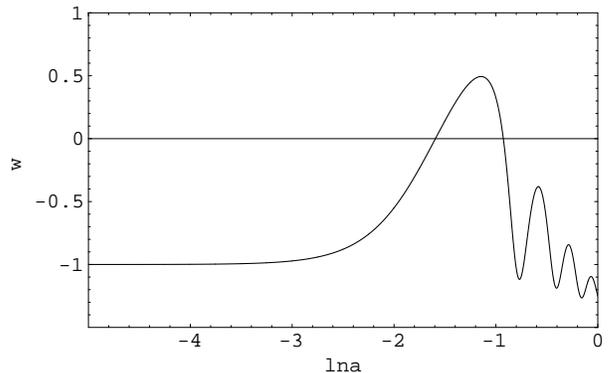}
\caption{ The state equation $w$ as a function of scale factor
$\ln{a}$ for the potential $V=
\frac{1}{2}m_{1}^{2}\phi^{2}+\frac{1}{2}m_{2}^{2}\sigma^{2}
+\lambda \phi^{2}\sigma^{2}$, where $ \frac{1}{2}m_{1}^{2}= 5.9
\times10^{-84}GeV^2, \frac{1}{2}m_{2}^{2}= 1.19\times
10^{-82}GeV^2 $, and $ \lambda=50 \times 10^{-120}$. }
\end{center}
\end{figure}

\section{Quintom model with Phantom field and neutrino}

The two-field models of Quintom dark energy considered above
consist of two scalar fields. In the following we study a model
where the Quintessence in (\ref{Ltot}) is replaced by the
neutrinos. In this model the neutrino will become a part of the
dark energy. There have been a lot of studies recently in the
literature on the coupled system of neutrinos interacting with the
dark energy scalars\cite{paper30, paper31, paper32, paper33,
paper34, paper35, paper36, paper37, paper38}, however in all of
these studies the scalar fields have canonical kinetic terms. In
the following we will show that a system with a Phantom and
neutrinos can naturally give rise to a realization of Quintom
models. Furthermore since the neutrinos are the particles existing
in the standard model of the elementary particle physics this
model introduces less degree of freedom in comparison to the
Quintom model considered above with two scalar fields.

The Lagrangian for such a coupled system is given\footnote{The
studies on the interacting Phantom dark energy with dark matter
are given in\cite{guo}} by
\begin{equation}
{\cal L}={\cal L}_\nu+{\cal L}_\phi+M(\phi)\bar{\nu}{\nu}\ ,
\label{Lnu}
\end{equation}
where  \begin{equation}
  {\cal L}_\phi=-\frac{1}{2}\partial_\mu
  \phi\partial^\mu\phi-V(\phi),
  \end{equation}
  with $V(\phi)$ the potential for Phantom field
and \begin{equation} {\cal L_{\nu}}=\bar{\nu}i\partial
\hspace{-0.2cm}/ \nu,
\end{equation}
which is the kinetic term of the neutrino. In (\ref{Lnu})
$M(\phi)\bar{\nu}{\nu}$ characterizes the interaction between the
Phantom and the neutrinos which for instance in Ref.\cite{paper33}
is given by \be\label{coupl} {\cal L}_{int}= e^{-\beta \frac{ \phi
}{M_{pl}}} \frac{2 }{f} l_{L}l_{L} H H+ h.c  , \ee where $\beta $
is the coefficient characterizing the strength of the Phantom
interaction with the neutrinos, $f$ is the scale of new physics
beyond the Standard Model
  which generates the $B-L$ violation, $l_{L},  H$ are the
  left-handed lepton and Higgs doublets respectively. When the
  Higgs field gets a vacuum expectation value $<H> \sim v$, the
  left-handed neutrino receives a Phantom field dependent Majorana mass
  $m_{\nu} \sim e^{-\beta \frac{ \phi
}{M_{pl}}} \frac{v^{2}}{f}$.

In general, with different $V(\phi)$ and $M(\phi)$ in (14) and (15) this
two-field Quintom model of dark energy behaves differently.
Firstly we will analyze the general feature of the cosmological
evolution for this class of models without specifying the explicit
form of $V(\phi)$ and $M(\phi)$. Due to the existence of the
interaction term $M(\phi)$, the evolution of the Phantom field is
determined by the effective potential which is the combination of
the potential $V(\phi)$ and\cite{paper33}
\begin{equation}
\bar{V}(\phi)=nM(\phi)\langle\frac{M(\phi)}{E}\rangle,
\end{equation}
with $n$ and $E$ being the number density and energy of the
neutrinos respectively and $\langle \rangle$ indicating the
thermal average. For relativistic neutrinos, the term
$\bar{V}(\phi)$ is greatly suppressed and the neutrinos and dark
energy decouple. For non-relativistic neutrinos, the effective
potential of the system is given by
$V_{eff}(\phi)=V(\phi)+nM(\phi)$. Consequently for the equation of
motion of the scalar field $\phi$ it is
\begin{equation}
\ddot{\phi}+3H\dot{\phi}-\frac{dV}{d\phi}-\frac{d\bar{V}}{d\phi}=0.
\end{equation}

The equation of state for such a coupled system is
\begin{equation}
w=\frac{-\frac{1}{2}\dot{\phi}^2-V(\phi)}{-\frac{1}{2}\dot{\phi}^2+
V(\phi)+nM(\phi)},
\end{equation}
which can be rewritten as
\begin{equation}
w=\frac{\Omega_{\phi}}{\Omega_{\phi}+\Omega_{\nu}}
w_{\phi}+\frac{\Omega_{\nu}}{\Omega_{\phi}+\Omega_{\nu}}w_{\nu}.
\end{equation}
From the equation above one can see that during the radiation
dominant period since $\Omega_{\nu}$ is much larger than
$\Omega_{\phi}$, $w$ of the coupled system will be around 1/3 and
in the matter dominant regime $w$ will be close to 0. However in
the late time when the Phantom energy dominant over neutrinos
$\Omega_{\phi}\gg\Omega_{\nu}$, $w$ gradually evolves into the
value smaller than -1. To illustrate this behavior in Fig. 4 we
plot the evolution of the equation of state $w$ of  a model where
$V=V_0 e^{- \lambda \frac{\phi}{M_{pl}}}$ and $M=\bar{M}e^{-
\gamma \frac{\phi}{M_{pl}}}$. In the numerical calculation we have
taken that $\lambda = 5.5$ and $\gamma = 2.5$. One can see from
this figure that $w$ changes from above -1 to below -1 as the
redshift decreases.

\begin{figure}[htbp]
\includegraphics[scale=0.8]{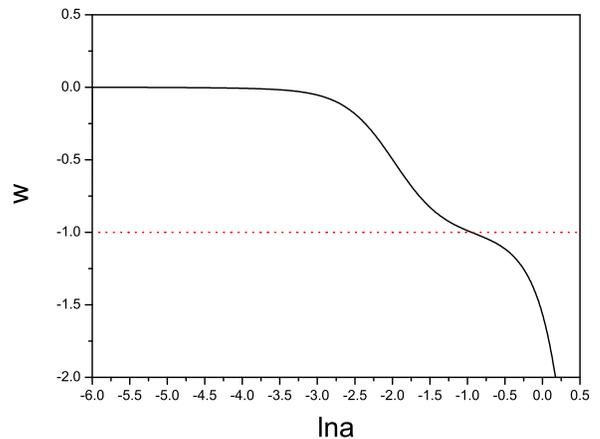}
\caption{Plot of the w of the system as a function of the scale
factor $\ln a$ for $V=V_0 e^{-5.5\frac{\phi}{M_{pl}}}$ and
$M=\bar{M}e^{-2.5\frac{\phi}{M_{pl}}}$. \label{Fig:spec}}
\end{figure}.

\section{Conclusion}

In this paper we have studied theoretically the possibility of
building the two-field dark energy model with an equation of state
across -1. In general within the framework of the general
relativity and the field theory, it is difficult to realize the
Quintom with a single scalar field model. Thus two fields are
required. In models with two scalar fields there will be a lot of
possibilities for the model building by choosing the potentials,
interactions and the model parameters. As shown in Figs. 2 and 3
the physics associated with this type of models and their
implications in cosmology are quite rich. In the model with
Phantom and neutrinos, as we emphasized in the paper since the
neutrinos are the particles existing in the standard model of the
elementary particle physics, compared with the model with two
scalar fields this model introduces less degree of freedom.
Furthermore the neutrino masses vary during the evolution of the
Phantom field, which makes this model more interesting. Before
concluding we should also point out that there exist possibilities
of building models of dark energy with equation of state across -1
beyond the field theory with minimal couplings to the gravity and
the four
 dimension\cite{LCCT}.

\section{Ackowlegments}
We thank   Bo Feng and   Xiaojun Bi for discussions. This work was
supported in part by National Natural Science Foundation of China
(grant Nos. 90303004, 19925523) and by Ministry of Science and
Technology of China( under Grant No. NKBRSF G19990754).


\begin{thebibliography}{99}

\bibitem{Riess} A. G. Riess et al., Astrophys. J. 607, 665 (2004).

\bibitem{Spergel} D. N. Spergel et al., Astrophys. J. Suppl. 148, 175 (2003).

\bibitem{Seljak} U. Seljak et al., astro-ph/0407372.

\bibitem{Quin} B. Ratra and P. J. E. Peebles, Phys. Rev. D 37, 3406 (1988).
C. Wetterich, Nucl. Phys. B 302, 668 (1988); I. Zlatev, L. Wang,
P. J. Steinhardt, Phys. Rev. Lett. 82 896(1999).

\bibitem{Caldw} R. R. Caldwell, Phys. Lett. B 545, 23 (2002).


\bibitem{Phantom} S. Nojiri and S. D. Odintsov, hep-th/0303117;
hep-th/0304131; hep-th/0306212; A. Feinstein and S. Jhingan,
hep-th/0304069; P. Singh, M. Sami and N. Dadhich, hep-th/0305110;
A. Yurov, astro-ph/0305019; P. F. Gonzalez-Diaz, astro-ph/0305559;
M. P. Dabrowski, T. Stachowiak and M. Szydlowski, hep-th/0307128;
H. Stefancic, astro-ph/0310904; P. F. Gonzalez-Diaz, Phys. Rev.
D68, 084016(2003);
 V. B. Johri, astro-ph/0311293; H. Stefancic,
astro-ph/0312484; M. Sami and A. Toporensky gr-qc/0312009; M.
Szydlowski, W. Czaja, and A. Krawiec, astro-ph/0401293; J. A. S.
Lima and J. S. Alcaniz, Phys.Lett. B600 191(2004); Z. K. Guo, Y.
S. Piao and Y. Z. Zhang, Phys. Lett. B594 247(2004); J. M.
Aguirregabiria, L. P. Chimento and R. Lazkoz astro-ph/0403157; F.
Piazza and S. Tsujikawa, JCAP 0407 004(2004); L. P. Chimento and
R. Lazkoz, gr-qc/0405020; astro-ph/0405518; astro-ph/0411068; E.
Elizalde, S. Nojiri and S. D. Odintsov, hep-th/0405034; V. K.
Onemli and R. P. Woodard, Phys. Rev. D70 107301(2004); P. F.
Gonzalez-Diaz and J. A. Jimenez-Madrid, Phys. Lett. B596 16(2004);
P. F. Gonzalez-Diaz and C. L. Siguenza, Nucl. Phys. B697
363(2004); S. Nojiri and S. D. Odintsov, hep-th/0408170; H.
Stefancic, astro-ph/0411630; I. Ya. Aref'eva, A. S. Koshelev and
S. Yu. Vernov, astro-ph/0412619; M. P. Dabrowski and T.
Stachowiak, hep-th/0411199; T. Brunier, V. K. Onemli and R. P.
Woodard, Class. Quant. Grav.22 59(2005); V. K. Onemli and R. P.
Woodard, Phys. Rev. D70 107301(2004); S. D. H. Hsu, A. Jenkins and
M. B. Wise, Phys. Lett. B597 270(2004).


\bibitem{stein} for recent analyses, see,
P. S. Corasaniti, M. Kunz, D. Parkinson, E. J. Copeland and  B. A.
Bassett, Phys. Rev. D70  083006(2004), astro-ph/0406608; S.
Hannestad and E. Mortsell
 JCAP 0409 001(2004), astro-ph/0407259; A. Upadhye, M. Ishak and P. J. Steinhardt,
astro-ph/0411803.




\bibitem{FWZ} B. Feng, X. Wang and X. Zhang, astro-ph/0404224, Phys. Lett.
B607, 35(2005).

\bibitem{FLPZ} B. Feng, M. Li, Y.Piao and X. Zhang,
astro-ph/0407432.

\bibitem{Piao} Y. S. Piao and E. Zhou, Phys. Rev. D68 083515(2003);
Y. S. Piao and Y. Z. Zhang, Phys. Rev. D70 063513(2004).

\bibitem{XBZ} J. Xia, B. Feng and X. Zhang, astro-ph/0411501.

\bibitem{Vikman} A. Vikman, astro-ph/0407107.

\bibitem{GPZZ} Z. Guo, Y. Piao, X. Zhang and Y. Zhang, astro-ph/0410654,
to appear in Phys Lett B.

\bibitem{WHU}  W. Hu, astro-ph/0410680; L. Perivolaropoulos, astro-ph/0412308; H. Wei and R. G.
Cai, hep-th/0501160;  S. Nojiri, S. D. Odintsov and S. Tsujikawa,
hep-th/0501025; R. R. Caldwell and M. Doran, astro-ph/0501104; Y.
Wei and Y. Tian, Class. Quant. Grav. 21:5347-5353(2004).


\bibitem{HW} I. P. C. Heard and D. Wands, Class.Quant.Grav. 19 5435(2002),
gr-qc/0206085.

\bibitem{GPZ} Z. K. Guo, Y. S. Piao and Y. Z. Zhang, Phys. Lett. B568 1(2003); Z. K. Guo, Y. S. Piao,
R. G. Cai and Y. Z. Zhang, Phys. Lett. B576 12(2003).

\bibitem{paper30}
 P. Q. Hung, hep-ph/0010126.


\bibitem{paper31}
 P. Gu, X. Wang and X. Zhang, Phys. Rev.
 D68, 087301 (2003); hep-ph/0307148.

\bibitem{paper32}
R. Fardon, A. E. Nelson and N. Weiner, astro-ph/0309800.

\bibitem{paper33}
X. Bi, P. Gu, X. Wang and X. Zhang, Phys. Rev.
 D69, 113007 (2004); hep-ph/0311022.

\bibitem{paper34}
P. Hung and H. Pas, astro-ph/0311131.


\bibitem{paper35}
 D. B. Kaplan, A. E. Nelson and N. Weiner,
 Phys. Rev. Lett. 93, 091801(2004); hep-ph/0401099

\bibitem{paper36}
R. D. Peccei, hep-ph/0404277; hep-ph/0411137 .

\bibitem{paper37}
H. Li, Z. Dai and X. Zhang, hep-ph/0411228.

\bibitem{paper38}
X. Zhang, hep-ph/0410292;
 E. I. Guendelman, A. B. Kaganovich,  hep-th/0411188.


\bibitem{guo}
Z. Guo, R. Cai and Y. Zhang, astro-ph/0412624,
 R. Cai and A. Wang, hep-th/0411025.



\bibitem{LCCT} B. Li, M. C. Chu, K. C. Cheung and A. Tang, astro-ph/0501367.

\end{thebibliography}
\end{document}